\documentstyle[epsfig]{elsart}
\newcommand{\gsim}{\mathrel{\rlap{\lower4pt\hbox{\hskip1pt$\sim$}}
\raise1pt\hbox{$>$}}}
\newcommand{\lsim}{\mathrel{\rlap{\lower4pt\hbox{\hskip1pt$\sim$}}
\raise1pt\hbox{$<$}}}
\newcommand{\sfrac}[2]{\mbox{\footnotesize $\frac{#1}{#2}$}}
\begin{document}
\begin{frontmatter}
%
%_______________________ Title, Authors ____________________________________
\title{Electromagnetic Pion Form Factor\\and Neutral Pion Decay Width}
\author{Craig D. Roberts}
\address{Physics Division, Bldg. 203, Argonne National Laboratory,\\
Argonne, Illinois 60439-4843}
%
%-------------------------------------------------------------------
\begin{abstract}
The electromagnetic pion form factor, $F_\pi(q^2)$, is calculated for
spacelike-$q^2$ in impulse approximation using a confining quark propagator,
$S$, and a dressed quark-photon vertex, $\Gamma_\mu$, obtained from
realistic, nonperturbative Dyson-Schwinger equation studies.  Good agreement
with the available data is obtained for $F_\pi(q^2)$ and other pion
observables, including the decay $\pi^0 \rightarrow \gamma\,\gamma$.  This
calculation suggests that soft, nonperturbative contributions dominate
$F_\pi(q^2)$ at presently accessible~$q^2$.
\end{abstract}
% \vspace*{\baselineskip}\\
%
% PACS: 13.40.Fn, 14.40.Aq, 12.38.Lg, 12.40.Aa
% See PRL Vol. 67 No. 25
% \vspace*{\baselineskip}\\
%
\end{frontmatter}

%..............................
\section{Introduction}
As a bound state of a light quark and antiquark, the pion is an ideal system
for exploring the application of different approaches to the study of bound
state structure in QCD, which is intrinsically nonperturbative.  Such studies
are constrained by the Goldstone boson nature of the pion.  The internal
structure of the pion affects its observable properties.  The pion
electromagnetic form factor, $F_\pi(q^2)$, is one observable that is
sensitive to this internal structure and it has been much
studied~\cite{CCP88,LAA89,IBG90}.

Perturbative QCD has been employed in order to estimate the behaviour of
$F_\pi(q^2)$ at large spacelike-$q^2$.  These analyses rely on the separation
of the amplitude into a product of soft and hard contributions using a
factorisation Ansatz~\cite{ER80LB80}, however, the applicability of this
approach to exclusive processes is uncertain.  In this factorisation approach
$F_\pi(q^2)$ is the product of a soft contribution that depends on the
bound-state Bethe-Salpeter amplitude, which provides only the overall
normalisation, and a hard contribution that is independent of the bound-state
Bethe-Salpeter amplitude and is taken to be given by the Born amplitude for a
collinear quark-antiquark pair, each massless, to scatter coherently from a
virtual photon.  This analysis yields
\mbox{$q^2\,F_\pi(q^2) = 16\,\alpha(q^2)\,f_\pi^2$} as
spacelike-$q^2\rightarrow\infty$, where $f_\pi\approx 92$~MeV and
$\alpha(q^2)$ is the running coupling constant in QCD.

It is not clear whether presently accessible values of $q^2$ are large enough
to test predictions based on perturbative analyses in QCD.  It is argued in
Ref.~\cite{ILS} that they are not; i.e., that the factorisation Ansatz is
invalid at presently accessible values of $q^2$ and hence that the $q^2$
dependence of the quark momentum distribution in the pion provides an
important contribution to $F_\pi(q^2)$.  This conclusion is supported by
Ref.~\cite{SW93} and by the fact that a good fit to the experimental data,
over the entire range of available $q^2$, is possible using the light-front
formulation of a relativistic constituent quark model~\cite{CCP88}, which has
no obvious connection with perturbative QCD.

\begin{figure}[h,t]
  \centering{\
     \epsfig{figure=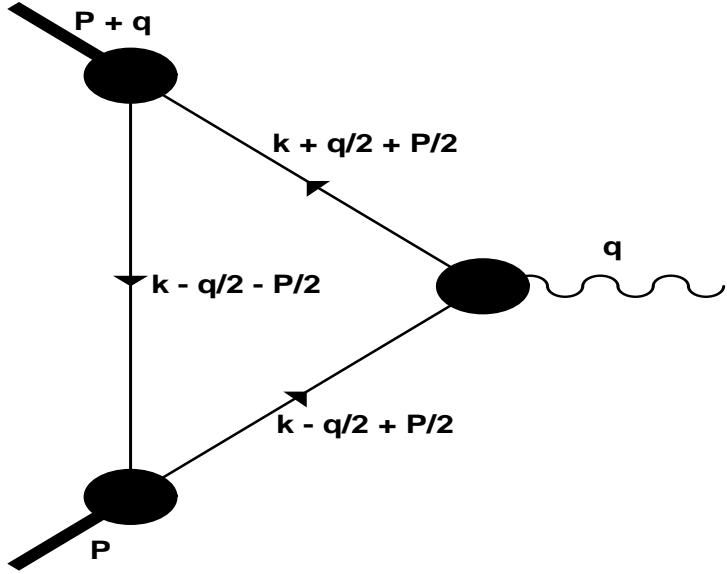,height=11cm,rheight=10cm,width=10cm}  }
\caption{A pictorial representation of the amplitude identified with the
$\pi$-$\pi$-$A_\mu$ vertex in impulse approximation: pions - thick, straight
external lines; $\langle\pi|\overline{q}q\rangle$ Bethe-Salpeter amplitudes -
filled circles at the $\pi$-legs; photon - wiggly line; dressed quark-photon
vertex, which satisfies the Ward-Takahashi Identity, Eq.~(\protect\ref{WTI})
- shaded circle at the the photon-leg; dressed quarks - thin internal
lines. \label{figppA}}
\end{figure}

Herein the impulse approximation to $F_\pi(q^2)$, illustrated in
Fig.~\ref{figppA}, is calculated for spacelike-$q^2$ as a phenomenological
application of the nonperturbative Dyson-Schwinger equation [DSE] approach to
QCD, which is reviewed in Ref.~\cite{DSErev}.  The primary elements of this
calculation are: 1) The dressed quark propagator, $S(p)$, which is confining
in the sense that it has no singularities that can lead to free-quark
production thresholds in Fig.~\ref{figppA}; i.e., there is {\it no quark
mass-shell}; 2) The pion Bethe-Salpeter amplitude, $\Gamma_\pi(p,P)$, which
is regular for spacelike values of $p$, the relative $q-\overline{q}$
momentum.  In the chiral limit, $m\to 0$, $\Gamma_\pi(p,P)$ is completely
determined by the dressed quark propagator, which is a manifestation of
Goldstone's theorem in DSE approach\cite{DS79}; and 3) The dressed
quark-photon vertex, $\Gamma_\mu(p_1,p_2)$, which is regular in the spacelike
region; i.e., away from resonances such as the $\rho$-meson, and follows from
extensive QED studies\cite{BC80,CPcoll,BR93}.  These properties, which
together are sufficient to ensure confinement, entail that this impulse
approximation calculation is free of both endpoint and pinch singularities,
which arise in perturbative analyses.

In phenomenological applications of the DSE approach the model dependence is
restricted to spacelike-$q^2\lsim 2$~GeV$^2$, and is realised in a modelling
of the form of the quark-quark interaction in the infrared.  This not only
incorporates information obtained about, for example, the gluon condensate in
the QCD sum rules approach~\cite{RWK92} but also extends it.  The calculation
of experimental observables in this approach therefore allows one to place
constraints on the qualitative and quantitative features of the effective
quark-quark interaction at small spacelike-$q^2$ in QCD and to infer the
$q^2$ scale where perturbative, model-independent, effects begin to dominate.

In Sec.~2 the impulse approximation and its primary elements [$S$,
$\Gamma_\pi$ and $\Gamma_\mu$] are discussed in detail.  The width
$\Gamma_{\pi^0\rightarrow\gamma\gamma}$ is calculated in impulse
approximation in Sec.~3 and shown to be {\it independent} of the details of
$S$, $\Gamma_\pi$ and $\Gamma_\mu$ in the chiral limit.  This illustrates the
manner in which anomalies are realised in the present framework.  The
calculation of $F_\pi(q^2)$ for spacelike-$q^2$, and other pion observables,
is described in Sec.~4 and the results compared with experiment.  The
behaviour of $F_\pi(q^2)$ at large spacelike-$q^2$ in impulse approximation
is determined analytically in Sec.~5: \mbox{$F_\pi(q^2)\propto 1/q^4$}, up to
$\ln[q^2]$-corrections.  This result is verified numerically and found to
become dominant only for spacelike-$q^2 \gsim 10$~GeV$^2$, which is presently
inaccessible experimentally.  This asymptotic form is a consequence of the
realisation of confinement explored herein.  Its validity or otherwise is not
an essential consequence of the DSE framework.  The results are summarised
and conclusions presented in Sec.~6.
%....................
\section{Impulse Approximation}
Herein all calculations are carried out in Euclidean space, with $\gamma_\mu$
hermitian and metric \mbox{$\delta_{\mu\nu}={\rm diag}(1,1,1,1)$}.

One may define the impulse approximation to the connected $\pi$-$\pi$-$A_\mu$
vertex in QCD as, with $m_u=m_d$,
\begin{eqnarray}
\label{LFpi}
\lefteqn{\Lambda_\mu(P+q,-P)= \frac{2 N_c}{f_\pi^2}\,
\int\sfrac{d^4k}{(2\pi)^4}\, {\rm tr}_D
\left[ \overline{\Gamma}_\pi(k;P+q) \times \right.} \\
& & \nonumber \left.
        S(k_{++})i\Gamma_\mu(k_{++},k_{-+})S(k_{-+})
        \Gamma_\pi(k-\sfrac{1}{2}q;-P) S(k_{--})\right]~,
\end{eqnarray}
where $q$ is the photon momentum and $P$ is the initial momentum of the pion.
Here the trace over colour and flavour indices has been evaluated leaving
only the trace over Dirac indices and
\begin{equation}
k_{\alpha\beta} = k + \sfrac{\alpha}{2} q + \sfrac{\beta}{2}P~.
\end{equation}
In Eq.~(\ref{LFpi}): $\Gamma_\mu(p_1,p_2)$ denotes the dressed quark-photon
vertex; $\Gamma_\pi(p;P)$ the pion Bethe-Salpeter amplitude, with $p$ the
relative momentum and $P$ the centre-of-mass momentum; and $S(p)$ the dressed
quark propagator.  The Bethe-Salpeter wave-function is
\begin{equation}
\label{BSEwf}
\chi_\pi(p;P) = S(p+\sfrac{1}{2}P)\, \Gamma_\pi(p;P) \, S(p-\sfrac{1}{2}P)~.
\end{equation}

The impulse approximation, Eq.~(\ref{LFpi}), is illustrated in
Fig.~\ref{figppA} and can be derived as an application of the formalism
described in Ref.~\cite{FT94}.  Its definition is only complete when the
functions $S$, $\Gamma_\pi$ and $\Gamma_\mu$ have been fully specified, which
is the subject of Secs.~\ref{quarkprop}, \ref{bethesalpeteramp} and
\ref{VtxAns}, respectively.

\subsection{Quark Propagator}
\label{quarkprop}
The dressed quark propagator in Eq.~(\ref{LFpi}) can be obtained by solving
the following Dyson-Schwinger equation [DSE]:
\begin{equation}
 S^{-1}(p) = i \gamma\cdot p + m
     + \sfrac{4}{3} g^2 \int \sfrac{d^4k}{(2\pi)^4} \gamma_\mu
         S(k) \Gamma_\nu^g (k,p) D_{\mu \nu}(p-k),     \label{fullDSE}
\end{equation}
where $m= (m_u+m_d)/2$ is the current-quark mass.  Here $D_{\mu\nu}(k)$ is
the dressed gluon propagator and $\Gamma_\mu^g(p_1,p_2)$ is the dressed
quark-gluon vertex; each of which satisfies its own DSE.  The general form of
the solution of Eq.~(\ref{fullDSE}) is
\begin{equation}
S(p) = - i \gamma\cdot p\, \sigma_V(p^2) + \sigma_S(p^2),
\end{equation}
which can also be written as
\begin{equation}
\label{ABform}
S(p) = \frac{1}{i\gamma\cdot p\, A(p^2) + m + B(p^2)}~.
\end{equation}
This equation has been much studied and the general properties of $\sigma_S$
and $\sigma_V$ in QCD are well known~\cite{DSErev}.  

As a simple example, in Ref.~\cite{BRW92} Eq.~(\ref{fullDSE}) was solved with
\begin{eqnarray}
\label{BRWMod}
g^2 D_{\mu\nu}(k) = \left(\delta_{\mu\nu} - \frac{k_\mu k_\nu}{k^2}\right)
                8 \pi^4 D \delta^4(k) \;\; & \; \mbox{and} \; &
\;\; \Gamma_\mu(p,p) = -i \partial_\mu S^{-1}(p)~,
\end{eqnarray}
where $D$ is a mass-scale parameter.  This Ansatz for the dressed gluon
propagator models the infrared behaviour of the quark-quark interaction in
QCD via an integrable infrared singularity, as suggested by
Refs.~\cite{Baker,Atkin,Pennington}, and and is sufficient to ensure
confinement, in the sense described below.  The Ansatz for the dressed-quark
gluon vertex is the result of extensive analysis of its general
form~\cite{BC80,CPcoll,BR93}.  The solution of Eq.~(\ref{fullDSE}) in
Ref.~\cite{BRW92} is
\begin{eqnarray}
\label{SS}
\lefteqn{\bar\sigma_S(y) = }\\
& & \nonumber
\frac{C}{2 \overline{m} y}\exp(-2y^2) J_1(4 \bar m y) 
+ \frac{\overline{m}^2}{y} 
        \int_0^\infty\,d\xi\, \xi\,K_1(\bar m \xi)\,J_1(y \xi)
        \exp\left(-\frac{\xi^2}{8}\right)~,
\end{eqnarray}
with $\overline{m} = m/\sqrt{2D}$, $y^2= p^2/(2D)$ and where $J_1$ and $K_1$
are Bessel functions, and
\begin{equation}
\label{SV}
\bar\sigma_V(y) = 
        \frac{1}{\bar m}\left( 
        \bar \sigma_S(y) + \frac{1}{4 y}\frac{d}{dy} \bar \sigma_S(y)\right)~,
\end{equation}
with $\bar\sigma_S(y^2) = \sqrt{2\,D}\,\sigma_S(k^2)$ and $\bar \sigma_V(y^2)
= 2\,D\,\sigma_V(k^2)$.  In Eq.~(\ref{SS}), $C$ is a parameter associated
with dynamical chiral symmetry breaking and it is not determined by
Eq.~(\ref{fullDSE}) with Eq.~(\ref{BRWMod}), while the integral, which is
associated with explicit chiral symmetry breaking, cannot be evaluated in
terms of known functions.

At large spacelike-$p^2$ one finds from Eqs.~(\ref{SS}) and (\ref{SV}) that 
\begin{eqnarray} 
\label{SUV}
\sigma_S(p^2) \approx \frac{m}{p^2} - \frac{m^3}{p^4} + \ldots \;\;
& \; \mbox{and} \; & \;\;
\sigma_V(p^2) \approx \frac{1}{p^2} - \frac{D+m^2}{p^4} + \ldots ~.
\end{eqnarray}
At large spacelike-$p^2$ in QCD one has at leading order
\begin{equation}
\sigma_S(p^2) \approx \frac{\hat{m}}{p^2
\left[\sfrac{1}{2}\ln\left(p^2/\Lambda_{\rm QCD}^2\right)\right]^d}
\end{equation}
with $\hat{m}$ a renormalisation point invariant and $d= 12/[33-2N_f]$; $N_f$
is the number of quark flavours.  One therefore sees that, neglecting
$\ln[p^2]$ terms, the model defined by Eqs.~(\ref{fullDSE}) and
(\ref{BRWMod}) incorporates asymptotic freedom.

Another feature of this model is that both $\sigma_S$ and $\sigma_V$ are
entire functions in the complex-$p^2$ plane with an essential singularity.
As a consequence the quark propagator does not have a Lehmann representation
and can be interpreted as describing a confined particle.  This is because,
when used in Eq.~(\ref{LFpi}), for example, this property ensures the absence
of free-quark production thresholds, under the reasonable assumptions that
$\Gamma_\pi$ is regular for spacelike-$p^2$ and $\Gamma_\mu$ is regular for
spacelike-$q^2$.  It follows from this that Eq.~(\ref{LFpi}) is free of
endpoint and pinch singularities.

This is a particular, sufficient manner in which to realise the requirement
that Fig.~\ref{LFpi} have no free-quark production thresholds, which is the
{\it definition} of confinement explored herein.  There are other, more
complicated, means of realising this definition but the effect is the
same~\cite{DSErev}.  Some of the phenomenological implications of a model
with a simple realisation of this confinement mechanism have been discussed
in Ref.~\cite{QCM}.

The solution described by Eqs.~(\ref{SS}) and (\ref{SV}) has a defect.  To
see this one sets $\bar m = 0$ in Eq.~(\ref{SS}), which yields
\begin{equation}
\sigma_S(y) = C e^{-2 y ^2}~.
\end{equation}
This poorly represents $\sigma_S$ for sufficiently large $y$ since, in the
absence of a bare mass and when chiral symmetry is dynamically broken, it is
known~\cite{HDP82} that
\begin{equation}
\label{PolSS}
\left.\sigma_S(y)\right|_{y\rightarrow \infty}
\rightarrow \, \frac{4 \pi^2 d}{3} \frac{\kappa}{y^2\,(\ln\,y^2)^{1-d}}
\end{equation}
with $\kappa = \,-\,(\ln[\mu^2/\Lambda_{QCD}^2])^{-d}\,\langle\bar q q
\rangle_{\mu^2}$, a renormalisation point invariant.  This defect results
from the fact that, although the form of $D_{\mu\nu}(k)$ in
Eq.~(\ref{BRWMod}) generates confinement, it underestimates the strength of
the coupling in QCD for sufficiently large $k^2$.  In the numerical studies
of Eq.~(\ref{fullDSE}) that have used a better approximation to
$D_{\mu\nu}(k)$~\cite{WKR91} there is no such defect.

\subsubsection{Approximate, algebraic quark propagator}
Herein, to avoid the need for a numerical solution of Eq.~(\ref{fullDSE}),
Eqs.~(\ref{SS}) and (\ref{SV}) are simply modified so as to restore the
missing strength at intermediate-$x\,(=y^2)$ and thereby provide a better
approximation to the realistic numerical solutions~\cite{WKR91}, while
retaining the confining characteristics present in the model example
described in association with Eqs.~(\ref{SS}) and (\ref{SV}).  

The following {\it approximating} algebraic forms are used:
\begin{eqnarray}
\label{SSM}
\lefteqn{\bar\sigma_S(x) = C e^{-2 x} +}\\
&&  
        \frac{1 - e^{- b_1 x}}{b_1 x}\,\frac{1 - e^{- b_3 x}}{b_3 x}\,
        \left( b_0 + b_2 \frac{1 - e^{- \Lambda x}}{\Lambda\,x}\right) 
        + \frac{\bar m}{x + \bar m^2}
                \left( 1 - e^{- 2\,(x + \bar m^2)} \right)~,
\nonumber
\end{eqnarray}
\begin{eqnarray}
\label{SVM}
\bar\sigma_V(x) & = & \frac{2 (x+\bar m^2) -1 
                + e^{-2 (x+\bar m^2)}}{2 (x+\bar m^2)^2}
                - \bar m C e^{-2 x},
\end{eqnarray}
which are entire functions, as in the model described above, and allow the
propagator to be consistent with realistic numerical solutions of
Eq.~(\ref{fullDSE}).

When $b_0 = 0 = b_2$, Eqs.~(\ref{SSM}) and (\ref{SVM}) provide an excellent
approximation to Eqs.~(\ref{SS}) and (\ref{SV})~\cite{MTRC94}, while for
nonzero values of these parameters it is clear that the behaviour given in
Eq.~(\ref{PolSS}) is recovered, up to $\ln[p^2]$-corrections.  These model
forms are also entire functions in the complex $p^2$ plane with an essential
singularity.

The expressions in Eqs.~(\ref{SSM}) and (\ref{SVM}) provide a six parameter
model of the quark propagator in QCD: $C$, $\bar m$, $b_0, \ldots, b_3$.
[$\Lambda (= 10^{-4})$ is introduced simply to decouple $b_2$ from the quark
condensate, as will be shown below.]  These parameters can be fitted to
experimental observables and Eq.~(\ref{fullDSE}) used to place constraints on
$D_{\mu\nu}(k)$, the effective gluon propagator.  In $S$ one therefore has an
implicit parametrisation of $D_{\mu\nu}(k)$ and hence a connection between
experimental observables and the nature of the effective quark-quark
interaction in the infrared.  A phenomenologically successful application of
this procedure may then make possible the use of precise experimental data as
a probe of the effective quark-quark interaction in the infrared.

%....................
\subsection{Pion Bethe-Salpeter Amplitude}
\label{bethesalpeteramp}
The Bethe-Salpeter amplitude in Eq.~(\ref{LFpi}) is the solution of the
homogeneous Bethe-Salpeter equation [BSE]:
\begin{equation}
\label{piBSE}
\Gamma^{rs}_\pi(p;P) = 
\int\,\sfrac{d^4k}{(2\pi)^4} \,K^{rs;tu}(k,p;P)\,
\left(S(k+\sfrac{1}{2}P)\Gamma_\pi(k;P)S(k-\sfrac{1}{2}P)\right)^{tu}
\end{equation}
where $P$ is the centre-of-mass momentum of the bound state, $p$ is the
relative momentum between the quarks in the bound state and the superscripts
are associated with the Dirac structure of the amplitude.  In the isospin
symmetric case, $m_u=m_d$, $K^{rs;tu}(p,k;P) \propto I_F$, the identity
matrix in flavour-space.  Further, since $\Gamma_\pi$ and $S(p)$ are $\propto
I_C$, the identity matrix in colour-space, then $K^{rs;tu}(p,k;P)\propto I_C$
also.

The {\it generalised}-ladder approximation is defined by the choice
\begin{equation}
\label{LaddK}
K^{rs;tu}(p,k;P) = \sfrac{4}{3}\, g^2 D_{\mu\nu}(p-k) \,
        (\gamma_\mu)^{rt}\,(\gamma_\nu)^{us}
\end{equation}
in Eq.~(\ref{piBSE}), with $D_{\mu\nu}(p-k)$ the {\it dressed} gluon
propagator and $S(p)$ the {\it dressed} quark propagator obtained from
Eq.~(\ref{fullDSE}).  This equation has been much studied~\cite{PCR89}.  In
using dressed quark and gluon propagators, the generalised-ladder
approximation is a significant advance over the ladder approximation familiar
in QED where perturbative fermion and photon propagators are used.  The
amplitude obtained as a solution in generalised-ladder approximation defines
the nonperturbative ``dressed-quark core'' of the meson and provides the
dominant contribution to physical observables, as will be seen herein.

The most general form of $\Gamma_\pi$ allowed by Lorentz covariance, which is
odd under parity transformations, is~\cite{LS69}
\begin{eqnarray}
\label{GGP}
\lefteqn{\Gamma_\pi(p;P) = i\gamma_5 \left\{\rule{0mm}{6mm}\,E(p;P) +
\right.}\\
& & \nonumber \left.\rule{0mm}{6mm}\
        i \gamma\cdot p \,p\cdot P \,F(p;P)
        + i \gamma\cdot P \,G(p;P) + [\gamma\cdot p ,\gamma\cdot P] \,H(p;P)
\right\}
\end{eqnarray}
where, since $\pi^0$ is even under charge-conjugation, $E$, $F$, $G$ and $H$
are even functions of $(p\cdot P)$.  The many studies of Eq.~(\ref{piBSE})
using Eq.~(\ref{LaddK})~\cite{PCR89} suggest that the dominant amplitude in
Eq.~(\ref{GGP}) is $E(p;P)$, with the other functions providing $\sim$ 10\%
contribution to physical observables; i.e., that it is a good approximation
to write
\begin{equation}
\label{GpiE}
\Gamma_\pi(p,P) = i\gamma_5 \, E(p;P)~.
\end{equation}

In generalised-ladder approximation $K$ is independent of $P$ and the
standard normalisation condition for $\Gamma_\pi$ reduces to the requirement
that, for \mbox{$P^2 = -m_\pi^2$}:
\begin{eqnarray}
\label{LAN}
P_\mu & = & N_c\,\int\,\sfrac{d^4k}{(2\pi)^4}\,{\rm tr}_D \left[ 
\overline{\Gamma}_\pi(k;P) \frac{\partial S(k_{0+})}{\partial P_\mu}
        \Gamma_\pi(k;-P) S(k_{0-}) \;+\right. \\
& &  \nonumber
\left.
\overline{\Gamma}_\pi(k;P) S(k_{0+})
        \Gamma_\pi(k;-P)\frac{\partial S(k_{0-}) }{\partial P_\mu} 
\right] ~.\nonumber
\end{eqnarray}

\subsubsection{Pion as Goldstone mode and quark-antiquark bound state}
In the chiral limit; i.e., when the current quark mass, $m$, is zero, the
pseudoscalar generalised-ladder approximation BSE and quark rainbow-DSE
[which has $\Gamma^g_\mu = \gamma_\mu$ in Eq.~(\ref{fullDSE})] are
identical~\cite{DS79}.  This entails that there is a massless excitation in
the pseudoscalar channel with
\begin{equation}
\label{gammapi0}
E(p;P^2=0) =\frac{1}{f_\pi}\, B_{m=0}(p^2)~,
\end{equation}
where $B(p^2)$ is given in Eq.~(\ref{ABform}) and $f_\pi$ is the calculated
normalisation constant.  This is the manner in which Goldstone's theorem is
realised in the Dyson-Schwinger equation framework.  In this case,
Eqs.~(\ref{SSM}) and (\ref{SVM}) completely determine $\Gamma_\pi$.

The dichotomy of the pion as both a Goldstone boson and a quark-antiquark
bound state is thus easily understood in the DSE approach.  One has dynamical
chiral symmetry breaking when, with $m=0$, Eq.~(\ref{fullDSE}) yields a
solution $B_{m=0}(p^2)\not\equiv 0$; i.e., when a momentum-dependent quark
mass is generated dynamically by the interaction of the quark with its own
gluon field.  For $m=0$ one also finds that the Bethe-Salpeter equation in
the pseudoscalar channel reduces to the quark DSE as $P^2\to 0$, where
$P_\mu$ is the total-momentum of the bound state.  It therefore follows,
without fine tuning, that if the quark-quark interaction is strong enough to
support dynamical chiral symmetry breaking then one has a massless,
pseudoscalar bound state of a strongly dressed quark and antiquark whose
Bethe-Salpeter amplitude is the quark mass function, Eq.~(\ref{gammapi0});
i.e., one has a zero mass bound state of a quark and antiquark, each of which
has an effective mass of $\sim 200$-$400$~MeV.  (The actual value depends on
the definition and the strength of the interaction.)

For $m \neq 0$, the pion Bethe-Salpeter amplitude must still vanish as the
relative momentum $p^2\rightarrow \infty$~\cite{VM90}.  A first, simple
approximation in this case is
\begin{equation}
\label{gammapi}
E(p;P^2=-m_\pi^2) \approx \frac{1}{f_\pi}\,B_{m\neq 0}(p^2)~,
\end{equation}
which, for small current-quark mass, is very good both pointwise and in terms
of the values obtained for physical observables~\cite{FR95}.  One notes that
using Eqs.~({\ref{SSM}) and (\ref{SVM}) in Eq.~(\ref{gammapi}) entails
\begin{equation}
\label{GUV}
\Gamma_\pi(p,P) \propto \frac{1}{p^2},
\end{equation}
which, up to the $\ln[p^2]$-corrections associated with the anomalous
dimension, reproduces the ultraviolet behaviour of the Bethe-Salpeter
amplitude given by QCD~\cite{VM90}. 
%.............................
\subsection{Quark-photon Vertex}
\label{VtxAns}
The quark-photon vertex, $\Gamma_\mu(p_1,p_2)$, satisfies a DSE that
describes both strong and electromagnetic dressing of the interaction.
Solving this equation is a difficult problem that has recently begun to be
addressed~\cite{MF93}. Much progress has been made in constraining the form
of $\Gamma_\mu(p_1,p_2)$ and developing a realistic
Ansatz~\cite{CPcoll,BR93}.

The bare vertex: $\Gamma_\mu(p_1,p_2) = \gamma_\mu$, is inadequate when the
fermion propagator has momentum dependent dressing because it violates the
Ward-Takahashi identity:
\begin{equation}
\label{WTI}
(p_1 - p_2)_\mu i \gamma_\mu \neq S^{-1}(p_1) - S^{-1}(p_2)~;
\end{equation}
and hence leads to an electromagnetic current for the pion that is not
conserved. 

An Ansatz~\cite{CPcoll} fulfilling the criteria~\cite{BR93} that it a)
satisfies the Ward-Takahashi identity; b) is free of kinematic singularities;
c) reduces to the bare vertex in the free field limit as prescribed by
perturbation theory; and d) has the same transformation properties as the
bare vertex under charge conjugation and Lorentz transformations, is
\begin{equation}
\label{genvertex}
\Gamma_\mu(p,k) = \Gamma_\mu^{\rm BC}(p,k) + \Gamma_\mu^{\rm T}(p,k)
\end{equation}
where~\cite{BC80}
\begin{eqnarray}
\label{VBC}
\lefteqn{\Gamma_{\mu}^{\rm BC}(p,k) =
\frac{\left[A(p^2) +A(k^2)\right]}{2}\;\gamma_{\mu} + }\\
 &  & \nonumber 
\frac{(p+k)_{\mu}}{p^2 -k^2}\left\{ \left[ A(p^2)-A(k^2)\right]
                 \frac{\left[ \gamma\cdot p + \gamma\cdot k\right]}{2}
- i\left[ B(p^2) - B(k^2)\right]\right\}  
\end{eqnarray}
and $(p-k)_\mu\,\Gamma_\mu^{\rm T}(p,k) = 0$ with $\Gamma_\mu^{\rm T}(p,p) =
0$.  $\Gamma_\mu^{\rm BC}$ in Eq.~(\ref{genvertex}) is {\it completely
determined} by the dressed quark propagator but $\Gamma_\mu^{\rm T}$ must be
determined otherwise.  Gauge covariance of the fermion propagator and
multiplicative renormalisability of the fermion DSE can be
used~\cite{CPcoll,BR93} to place constraints on $\Gamma_\mu^{\rm T}(p,k)$.
Using the bare quark propagator, which has $A=1$ and $B=$~constant,
$\Gamma_\mu^{\rm T}=0$.

The definition of the impulse approximation is complete with 
\begin{equation}
\label{GTZ}
\Gamma_\mu(p,k) \approx \Gamma_\mu^{\rm BC}(p,k)~
\end{equation}
in Eq.~(\ref{LFpi}).

\subsubsection{Vector meson dominance}
The electromagnetic pion form factor, $F_\pi(q^2)$, is an analytic function
but for a cut extending from $q^2= -\infty$ to $-4m_\pi^2$; i.e., on the
timelike real axis.  It is real for $q^2>-4m_\pi^2$.  Generalised vector
meson dominance is the observation that the pion form factor is completely
determined by the spectral density associated with the $J^{PC}=1^{--}$
pion-photon vertex, which is the discontinuity across this cut.

Models of vector meson dominance consist in a phenomenological Ansatz for
this spectral density.  A common model is a single, $\rho$-meson pole
representation.  Since the vertex in Equation~(\ref{GTZ}) is free of
kinematic singularities, by construction in accordance with criterion b)
above, vector meson pole contributions of this type are {\it explicitly
excluded} in impulse approximation.

Explicit vector-meson--photon mixing contributions to $F_\pi(q^2)$ enter
through the dressed quark-photon vertex and only appear in
\mbox{$\Gamma_\mu^{\rm T}(p,k)$} because vector mesons are resonances, which
are only uniquely defined on shell where they are transverse.  They appear
via resonance contributions in the timelike region to the hadronic component
of the photon vacuum polarisation.

For $q^2= 0$ the differential form of the Ward identity,
$\Gamma_\mu(p,p)=-i\partial_\mu S(p)$, completely determines the quark-photon
vertex ($\Gamma_\mu^{\rm T}(p,p)=0$).  This entails that the pion charge
radius is primarily determined by the dressed quark propagator and is not
very sensitive to $\Gamma_\mu^{\rm T}(p,k)$.  In the sense of generalised
vector meson dominance described above, this admits an interpretation that
the charge radius is primarily determined by the non-resonant part of the
full spectral density near threshold.  To go further and identify this as the
``tail of the $\rho$-meson resonance'' involves additional model assumptions
since the vector mesons are resonances, whose definition off-shell is
arbitrary.

This point is also discussed in Ref.~\cite{ABR95}.

\subsection{Current Conservation}
Using Eq.~(\ref{GTZ}) and the identities: 
\begin{equation}
\begin{array}{ll}
 S(-k)^T = C^\dagger\,S(k)\,C ;  & 
 \Gamma_\pi^T(-k;p)  =   C^\dagger\,\Gamma_\pi(k;p)\,C ;\\
 \bar\Gamma_\pi^T(-k;p)  =   C^\dagger\,\bar\Gamma_\pi(k;p)\,C ; &
 \Gamma_\mu^T(-k,-p)  =  - C^\dagger\,\Gamma_\mu(p,k)\,C , 
\end{array}
\end{equation}
where \mbox{$C=\gamma_2\gamma_4$} is the charge conjugation matrix, one finds
easily that the $\pi$-current is conserved:
\begin{equation}
q_\mu \Lambda_\mu(P+q,-P) = 0~.
\end{equation}

For elastic scattering, with $[2 q\cdot P + q^2=0]$, one can therefore write
\begin{equation}
\label{LamFQ}
\Lambda_\mu(P+q,-P) = \left( 2 P+ q\right)_\mu F_\pi(q^2,P^2)~.
\end{equation}

One obtains similarly that
\begin{eqnarray}
\lefteqn{\Lambda_\mu(P,-P) = 2 P_\mu\,F_\pi(0,P^2) = 
2 N_c \int\sfrac{d^4 k}{(2\pi)^4} 
{\rm tr}_D \left[ 
\overline{\Gamma}_\pi(k;P) \frac{\partial S(k_{0+}) }{\partial P_\mu}
\times\right. }\\
& &  \nonumber \left.\Gamma_\pi(k;-P) S(k_{0-})
 +\overline{\Gamma}_\pi(k;P) S(k_{0+})
        \Gamma_\pi(k;-P) \frac{\partial S(k_{0-}) }{\partial P_\mu}
\right]~. \nonumber
\end{eqnarray}
Comparing this with Eq.~(\ref{LAN}) it is clear that in impulse approximation
$F_\pi(0,P^2)=1$ only if the Bethe-Salpeter kernel is independent of $P$;
i.e., impulse approximation combined with a $P$-independent Bethe-Salpeter
kernel provides a consistent approximation scheme.  In this case one has, in
the chiral limit \mbox{$P^2 = -m_\pi^2 = 0$}:
\begin{eqnarray}
\lefteqn{f_{\pi}^2  =\frac{N_c}{8\pi^2}\int_{0}^\infty\,ds\,s\,B(s)^2\,
 }\label{Fpi}\\
& & \left(  \sigma_{V}^2 - 
2 \left[\sigma_S\sigma_S' + s \sigma_{V}\sigma_{V}'\right]
- s \left[\sigma_S\sigma_S''- \left(\sigma_S'\right)^2\right]
- s^2 \left[\sigma_V\sigma_V''- \left(\sigma_V'\right)^2\right]\right)~,
\nonumber
\end{eqnarray}
with $s=p^2$. 

This shows that the impulse approximation to the form factor,
Eq.~(\ref{LFpi}), is regular in the chiral limit.  In fact, the calculated
results are only weakly dependent on $m_\pi$.  The study of Ref.~\cite{ABR95}
indicates that, at $m_\pi=0.14$~GeV, Eq.~(\ref{LFpi}) provides the dominant
contribution to $F_\pi(q^2)$ for spacelike-$q^2$ and that pion-loops are
unimportant.
%....................
\section{Neutral pion decay}
\label{secpigg}
In Euclidean space the matrix element for the decay $\pi^0\rightarrow
\gamma\gamma$ can be written
\begin{equation}
\label{Mpi0}
{\cal M}(k_1,k_2)= -2\,i\, \frac{\alpha_{em}}{\pi f_\pi}\,
        \epsilon_{\mu\nu\rho\sigma}
        \epsilon_\mu(k_1)\,\epsilon_\nu(k_2)\,k_{1\rho}\,k_{2\sigma}\,
                G(k_1\cdot k_2)~,
\end{equation}
where $k_i$ are the photon momenta and $\epsilon(k_i)$ are their polarisation
vectors.  Here, the $\pi^0$ momentum is $P=(k_1 + k_2)$ and $P^2=2\,k_1\cdot
k_2$.  

Using Eq.~(\ref{Mpi0}) one finds easily that
\begin{equation}
\Gamma_{\pi^0\rightarrow \gamma\gamma} = 
        \frac{m_\pi^3}{16\pi}\,\left(\frac{\alpha_{em}}{\pi f_\pi} \right)^2
        G(-m_\pi^2)^2~.
\end{equation}
Experimentally one has \mbox{$
\Gamma_{\pi^0\rightarrow \gamma\gamma} = (7.74 \pm 0.56)~\mbox{eV}
$}, which corresponds to 
\begin{equation}
g_{\pi^0\gamma\gamma} \equiv G(-m_\pi^2) = 0.500 \pm 0.018~,
\end{equation}
using $m_{\pi^0}= 135$~MeV and $f_\pi=92.4$~MeV. 

In impulse approximation one has
\begin{eqnarray}
\label{gpigg}
\lefteqn{i\frac{\alpha_{em}}{\pi f_\pi}\,
        \epsilon_{\mu\nu\rho\sigma}\,k_{1\rho} \,k_{2\sigma} 
        \,G(k_1\cdot k_2) = \sfrac{1}{4\pi^2}\,
\int\,d^4k\,{\rm tr}_D\left[ \rule{0mm}{5mm} S(k)\,\right.}\\
& & \nonumber \left.\rule{0mm}{5mm}
i\Gamma_\mu(k,k-k_1)\,S(k-k_1)\,
        \Gamma_\pi(k+k_2/2;P)\,S(k+k_2) \,i\Gamma_\nu(k+k_2,k)\,\right]~.
\nonumber
\end{eqnarray}

In the chiral limit, $P^2=0$, and using Eq.~(\ref{GpiE}), one easily obtains
\begin{eqnarray}
\label{G0}
\lefteqn{g_{\pi^0\gamma\gamma}^0 \equiv G(0) =
\int_0^\infty\,ds\,s\,[f_\pi\,E(s;0)]\,A\,\sigma_V\,\times
 } \\
& & 
\left\{\rule{0mm}{5mm} A\,\left[\sigma_V\,\sigma_S 
        + s \left(\sigma_V'\,\sigma_S - \sigma_V\,\sigma_S'\right)\right]
+ s\,\sigma_V\,\left(A'\,\sigma_S - B'\sigma_V\right)\right\}~.
\nonumber
\end{eqnarray}
Now, defining
\begin{equation}
C(s) = \frac{B(s)^2}{s\,A(s)^2} = \frac{\sigma_S(s)^2}{s\,\sigma_V(s)^2}
\end{equation}
one obtains a dramatic simplification and Eq.~(\ref{G0}) becomes
\begin{equation}
g_{\pi^0\gamma\gamma}^0 = -\,\int_0^\infty\,ds\,
\frac{f_\pi\,E(s;0)}{B(s)}\,
        \frac{C'(s)}{[1+C(s)]^3}~.
\end{equation}

It therefore follows from the manner in which dynamical chiral symmetry
breaking is realised in the Dyson-Schwinger equation framework; i.e., from
Eq.~(\ref{gammapi0}), that in the chiral limit
\begin{equation}
\label{gpigg0}
g_{\pi^0\gamma\gamma}^0 = \int_0^\infty\,dC\,\frac{1}{(1+C)^3} = \frac{1}{2}~,
\end{equation}
since $C(s=0)=\infty$ and $C(s=\infty)=0$.  Hence, the experimental value is
reproduced {\it independent} of the details of $S(p)$.  

This illustrates the manner in which the Abelian anomaly is incorporated in
the Dyson-Schwinger equation framework.  Similar results are obtained for the
Wess-Zumino five-pseudoscalar term~\cite{PRC88} and $\gamma\pi\pi\pi$
interaction~\cite{AR95}.

In order to obtain the result in Eq.~(\ref{gpigg0}) it is essential that, in
addition to Eq.~(\ref{gammapi0}), the photon-quark vertex satisfy the Ward
identity.  This is not surprising.  However, the fact that one must dress all
of the elements in a calculation consistently is often overlooked.  

The subtle cancellations that are required to obtain this result also make it
clear that it cannot be obtained in model calculations where an arbitrary
cutoff function (or ``form-factor'') is introduced into each integral.  The
fact that $E(p;P^2=0)$ is the pion Bethe-Salpeter amplitude and
$f_\pi\,E(p;P^2=0)= B(p^2)$ in the chiral limit, is crucial.

A good approximation to $g_{\pi^0\gamma\gamma}$ for nonzero pion mass is
given by Eq.~(\ref{gpigg}) with Eq.~(\ref{gammapi}).

\section{Calculated Spacelike Form Factor}
The impulse approximation to $F_\pi(q^2)$ is defined by Eq.~(\ref{LFpi}) with
the dressed quark propagator obtained from Eq.~(\ref{fullDSE}), the
Bethe-Salpeter amplitude obtained using Eqs.~(\ref{piBSE}), (\ref{LaddK}) and
the dressed quark-photon vertex in Eq.~(\ref{GTZ}).

If, at large spacelike-$q^2$, each of the elements in Eq.~(\ref{LFpi})
behaves as prescribed by the renormalisation group in QCD then this amplitude
reduces to that studied in perturbative analyses of the elastic form factor.
[To see that the ``hard-scattering'' contribution is included one need only
once-iterate the $\pi$-$q$-$\overline{q}$ vertices using Eq.~(\ref{piBSE}).]
It is therefore plausible that a model such as the one described by
Eqs.~(\ref{SSM}), (\ref{SVM}) and (\ref{gammapi}), constructed so as to
preserve this asymptotic behaviour, should provide a result for the large
spacelike-$q^2$ behaviour of $F_\pi(q^2)$ whose only model dependence is in
the confinement mechanism.

Such a model will provide an extrapolation to small spacelike-$q^2$ that is
sensitive to the model parameters.  The DSE framework provides an
interpretation of these model parameters in terms of qualitative and
quantitative features of the quark-quark interaction in the infrared.

To evaluate Eq.~(\ref{LFpi}) it is convenient to work in the Breit frame with
\begin{eqnarray}
q=(0,0,-q,0) & \;\; {\rm and}\;\; & 
P=(0,0,\sfrac{1}{2}q,i\sqrt{m_\pi^2 + \sfrac{1}{4}q^2}\,)
\end{eqnarray}
in which case, with 
\mbox{$k = k\,(\sin\beta\,\sin\theta\,\cos\phi,
        \sin\beta\,\sin\theta\,\sin\phi,
        \sin\beta\,\cos\theta,\cos\beta)$},
\begin{eqnarray}
k\cdot q & = & -k \, q\, \sin\beta\,\cos\theta\, \\
k\cdot P & = & \sfrac{1}{2} k \,q \sin\beta\,\cos\theta + 
                i k \sqrt{m_\pi^2 + \sfrac{1}{4}q^2} \, \cos\beta
\end{eqnarray}
and one is left with three integrals to evaluate, one radial and two angular.

The definition of confinement employed herein ensures that, for all
spacelike-$q^2$, the integrand is regular and hence that the integrals can be
evaluated using straightforward Gaussian quadrature techniques; i.e., there
are no endpoint or pinch singularities.

\subsection{Fitting the Parameters}
Equations~(\ref{SSM}), (\ref{SVM}) and (\ref{gammapi}) provide a six parameter
model [$C$, $\overline{m}$, $b_0\ldots b_3$] of the nonperturbative
dressed-quark substructure of the pion based on DSE studies.  These
parameters are fixed herein by requiring that the model reproduce, as well as
possible, the following experimental values of the dimensionless quantities:
\begin{equation}
\begin{array}{ccc}
\displaystyle
\frac{f_\pi}
{\langle \overline{q}q\rangle^{\frac{1}{3}}} = 0.423~, &\;\;
f_\pi\,r_\pi = 0.318~,\;\;
\displaystyle
\frac{m_\pi^2}
{\langle \overline{q}q\rangle^{\frac{2}{3}}} = 0.396~;
\end{array}
\end{equation}
the dimensionless $\pi$-$\pi$ scattering lengths (see
Refs.~\cite{RCSI94,Poc95} for a discussion):
\begin{equation}
\begin{array}{cc}
a_0^0 = 0.26 \pm 0.05~,  & a_0^2 = -0.028 \pm 0.012~, \\
a_1^1 = 0.038 \pm 0.002~, & a_2^0 = 0.0017 \pm 0.0003~;
\end{array}
\end{equation}
and a least-squares fit to $F_\pi(q^2)$ on the spacelike-$q^2$ domain:
$[0,4]$ GeV$^2$.  The fitting procedure was performed using the expression
for the pion decay constant, $f_\pi$, given in Eq.~(\ref{Fpi}),
\begin{eqnarray}
\label{cndst}
\lefteqn{\langle\overline{q}q\rangle_{\mu^2} = }\\
& & \nonumber\,-\,
\left(\ln\frac{\mu^2}{\Lambda_{QCD}^2}\right)^\alpha
\lim_{\Lambda_{UV}^2\rightarrow \infty}
\left(\ln\frac{\Lambda_{UV}^2}{\Lambda_{QCD}^2}\right)^{-\alpha}
\frac{3}{4\pi^2}\int_0^{\Lambda_{UV}^2}\,ds\,s\left( \sigma_S(s)
        - \frac{m}{s+m^2}\right)~,
\end{eqnarray}
$s=p^2$ and $\Lambda_{QCD} = 0.20$~GeV, following Ref.~\cite{WKR91} but with
$\alpha = 1$ rather than the anomalous dimension, $d$, because the associated
$\ln[p^2]$-corrections have not been included in Eqs.~(\ref{SSM}) and
(\ref{SVM}), and the expressions for $a_0^0$, $a_0^2$, $a_1^1$, $a_2^0$ and
$r_\pi$ given in Ref.~\cite{RCSI94}.  Using Eq.~({\ref{SSM}) in
Eq.~(\ref{cndst}) one obtains
\begin{equation}
\langle\overline{q}q\rangle_{\mu^2} = \,-\, (2\,D)^{\frac{3}{2}}\,
\left(\ln\frac{\mu^2}{\Lambda_{QCD}^2}\right) \,
\frac{3}{4\pi^2}\,\frac{b_0}{b_1\,b_3}~,
\end{equation}
which is independent of $b_2$ for arbitrarily small but nonzero $\Lambda$.

Following this procedure one obtains
\begin{eqnarray}
& & \begin{array}{cc}
C = 0.0422~, & \overline{m} = 0.0111~,
\end{array} \nonumber \\
& & \begin{array}{cccc}
b_0 = 0.135~, & b_1 = 2.48~, & b_2 = 0.502 ~,& b_3 = 0.168~.
\end{array}
\label{ParamV}
\end{eqnarray}
The mass scale is set by requiring equality between the percentage error in
$f_\pi$ and $r_\pi$, which yields $D=0.132$~GeV$^2$.  

The results obtained are robust in the sense that they are not sensitive to
the functional form of the quark propagator.  Different functional forms,
provided they ensure confinement in the manner described herein, whose
parameters are allowed to vary in order to provide a good fit to pion
observables, yield a numerical form for the quark propagator that is,
pointwise, approximately the same on the integration domain explored by pion
observables.  Since the quark propagator implicitly constrains the gluon
propagator via Eq.~(\ref{fullDSE}), the robust nature of the study makes it
plausible that pion observables can be used to constrain the dressed-gluon
propagator.

To make the connection with the gluon propagator explicit is computationally
more intensive, with a calculation of the pion form factor requiring a
solution of Eq.~(\ref{fullDSE}) off the real-spacelike axis.  A first step in
that direction is made in Ref.~\cite{FR95}, in which a one-parameter model
dressed gluon propagator is employed in the calculation of all the
observables considered herein, except for the form factor.  That study
illustrates that all of the parameters in the quark propagator are correlated
via the single parameter in the gluon propagator and that one can fit the
available pion data to the same level of agreement with only one parameter,
in addition to the current quark mass.  The quark propagator {\it determined}
by this gluon propagator agreed well in form and magnitude with the
parametrisation employed herein evaluated with the best-fit parameters.

\subsection{Results for $F_\pi(q^2)$ and Other Observables.}
\label{secivb}
In Table~\ref{tabres} the low-energy physical observables calculated with the
parameter set of Eq.~(\ref{ParamV}) are compared with their experimental
values.  In this table \mbox{$m^{\rm ave}=\left( m_u +m_d\right)/2$ } and
$m_\pi$ is obtained from: \mbox{$m_\pi^2\,f_\pi^2 = -\,2\, m^{\rm
ave}_{\mu^2}\, \langle \bar q q \rangle_{\mu^2}$}.  Each of the calculated
quantities tabulated here was evaluated at the listed value of $m_\pi$; i.e.,
the chiral limit expressions for $f_\pi$, Eq.~(\protect\ref{Fpi}), $r_\pi$,
Eq.~(A21) of Ref.~\protect\cite{RCSI94}, and $g_{\pi^0\gamma\gamma}$,
Eq.~(\protect\ref{G0}), were not used, however, the finite pion mass
corrections are less than 1\% for each of these.  The experimental values of
$f_\pi$, $m^{\rm ave}$, $m_\pi$ and $g_{\pi\gamma\gamma}$ are extracted from
Ref.~\protect\cite{PDG}; $r_\pi$ from Ref.~\protect\cite{Exp86}; and the
scattering lengths are discussed in Ref.~\protect\cite{RCSI94,Poc95}.  The
value of \mbox{$\langle \bar q q\rangle$} is that typically used in QCD sum
rules analysis and the fitting error allows for deviations of 50\%.  The
results presented in the table support the notion that, in an expansion of
the pion mass in powers of the current-quark mass, the leading order term
dominates. The results presented herein are representative of DSE
studies~\cite{PCR89,FR95}.

\begin{table}[h,t]
\caption{A comparison between the low-energy $\pi$ observables calculated
 using the parameters of Eq.~(\protect\ref{ParamV}) and their experimental
 values (see text for sources) .\label{tabres} }
\begin{center}
\begin{tabular}{|cll|} \hline
   & Calculated  & Experiment  \\ \hline
  $f_{\pi} \; $    &  ~0.0837 GeV &   ~0.0924 $\pm$ 0.001     \\ \hline
  $-\langle \bar q q \rangle^{\frac{1}{3}}_{1\,{\rm GeV}^2}$ & ~0.221 &  
        ~0.220 $\pm$ 0.050\\ \hline
  $m^{\rm ave}_{1\,{\rm GeV}^2}$ & ~0.0057 & ~0.008 $\pm$ 0.007
                \\ \hline
 $m_{\pi} \; $    & ~0.132  & ~0.138  \\ \hline\hline
 $r_\pi \;$ & ~0.595 fm & ~0.663$\pm$ 0.006  \\  \hline\hline
 $g_{\pi^0\gamma\gamma}\;$ & ~0.498 (dimensionless) & ~0.500 $\pm$ 0.018\\  
                        \hline\hline
 $a_0^0 \;  $ & ~0.191  & ~0.26$\pm$0.05 \\ \hline
 $a_0^2 \;  $ & -0.0543 & -0.028 $\pm$ 0.012 \\ \hline
 $2 a_0^0-5 a_0^2\;$ & ~0.654 & ~0.66 $\pm$ 0.12 \\ \hline
 $a_1^1 \;  $ & ~0.0380 & ~0.038 $\pm$ 0.002\\ \hline
 $a_2^0 \; $  & ~0.00170 & ~0.0017 $\pm$ 0.0003\\ \hline
 $a_2^2 \;  $ & -0.000286 & \\ \hline
\end{tabular}
\end{center}
\end{table}

In Fig.~\ref{pipwa} the five $\pi$-$\pi$ partial wave amplitudes associated
with the scattering lengths in the table, calculated using the formulae in
Ref.~\cite{RCSI94}, which do not take final-state $\pi$-$\pi$ interactions
into account, are plotted.  They are in reasonable agreement with the data up
to $x \approx 3$, which corresponds to $E \approx 4\,m_\pi$.  
\begin{figure}[h,t]
  \centering{\
     \epsfig{figure=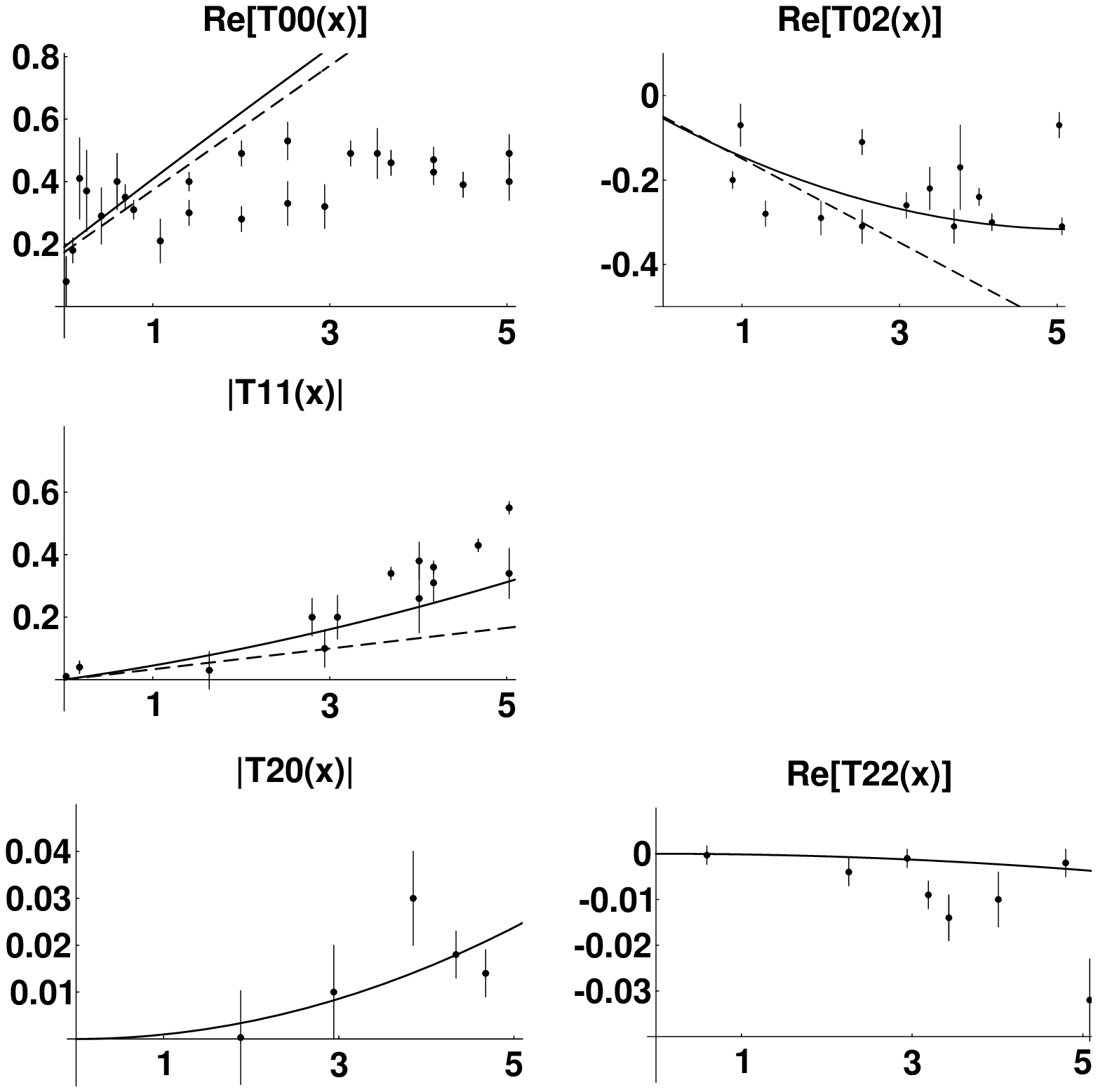,height=13cm,rheight=11cm,width=11cm}  }
\caption{A comparison between the $\pi$-$\pi$ partial wave amplitudes
calculated using Eq.~(\protect\ref{ParamV}) (solid line) and experiment (the
data are taken from Fig.~4 in Ref.~\protect\cite{RCSI94}).  The dashed line
is the current algebra prediction of Ref.~\protect\cite{SW66}. This is
identically zero for $T_2^0(x)$ and $T_2^2(x)$.  The dimensionless variable
$x= E^2/(4 m_\pi^2) - 1$ is defined so that threshold is at $x=0$.  Note that
$x=3$ corresponds to $E= 4 m_\pi$.\label{pipwa}}
\end{figure}

The form factor, $F_\pi(q^2)$, at small spacelike-$q^2$ is shown in
Fig.~\ref{smq} and for larger spacelike-$q^2$ in Fig.~\ref{lgq}.  Given that
the ``experimental'' point at $q^2=6.3$~GeV$^2$, measured in pion
electroproduction~\cite{Exp78}, depends strongly on the model used to separate
strong and electromagnetic effects, the model agrees well with the
experimental data.

\begin{figure}[h,t]
  \centering{\
     \epsfig{figure=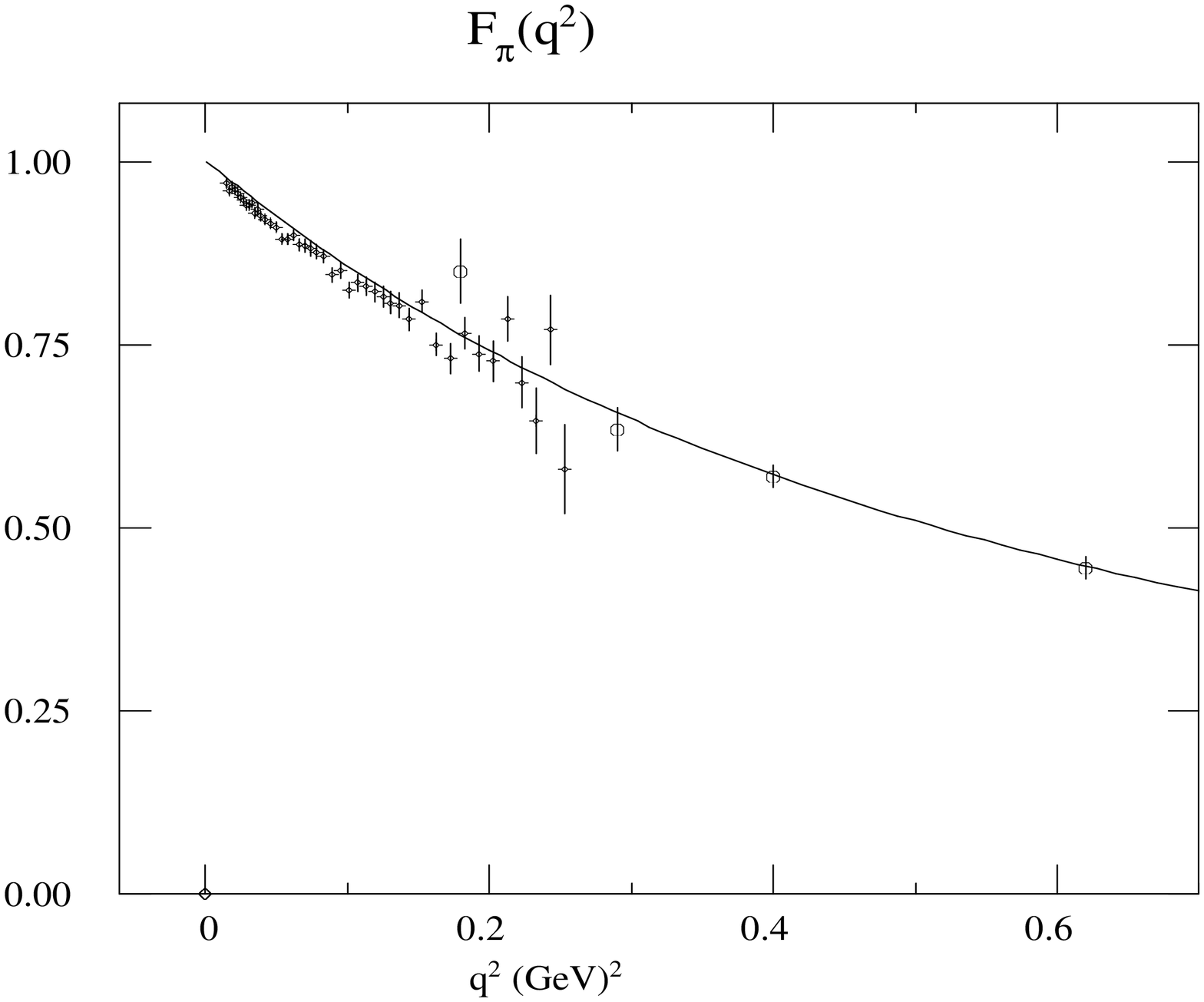,height=11cm,rheight=11cm,width=13cm,angle=-90}  }
\caption{The impulse approximation to $F_\pi(q^2)$ calculated using the
parameters in Eq.~(\protect\ref{ParamV}). The experimental data are from
Refs.~\protect\cite{Exp76} (circles) and \protect\cite{Exp86}
(crosses).\label{smq}}
\end{figure}

\begin{figure}[h,t]
  \centering{\
     \epsfig{figure=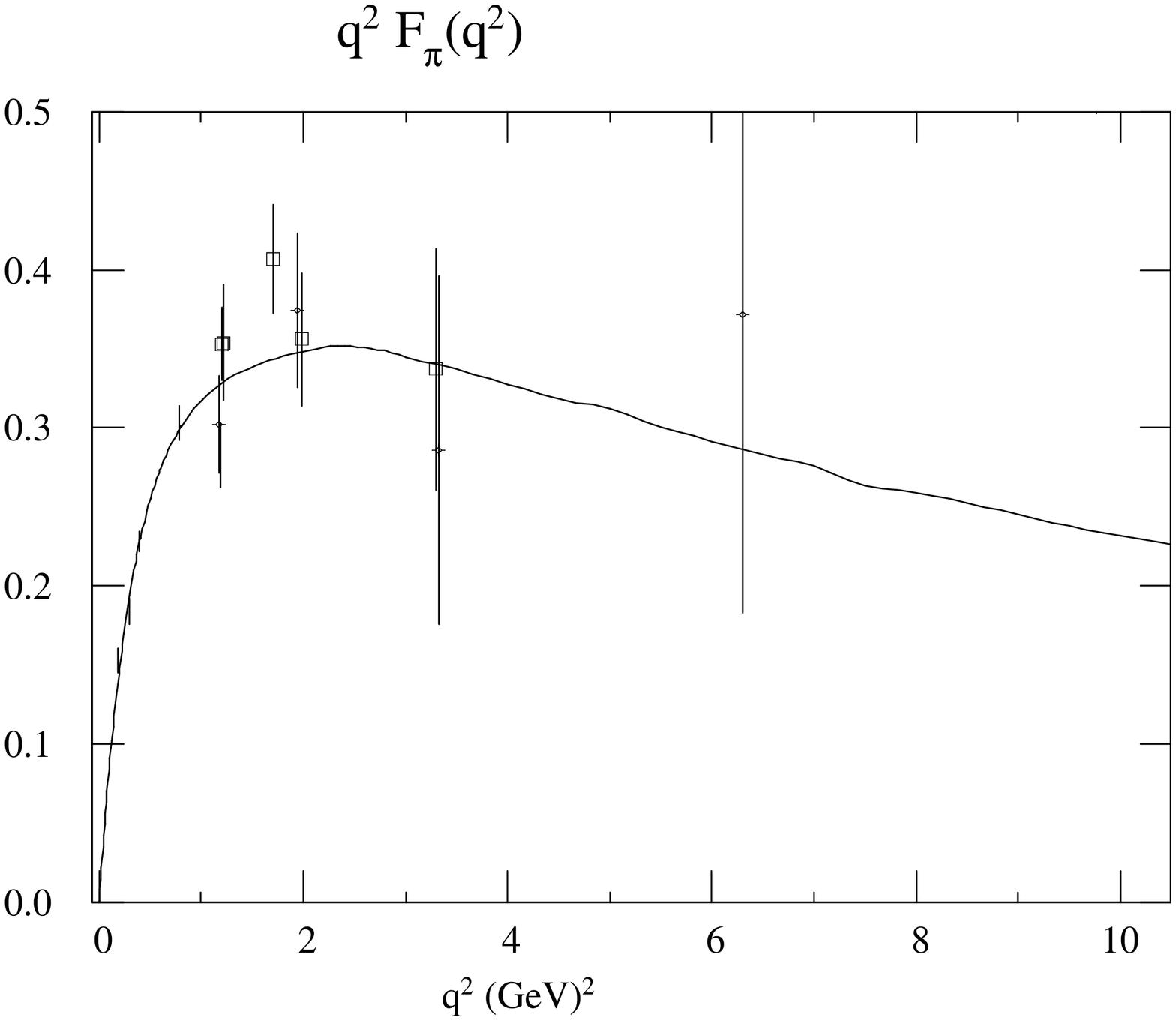,height=11cm,rheight=11cm,width=13cm,angle=-90}  }
\caption{The impulse approximation to $q^2\,F_\pi(q^2)$ calculated using the
parameters in Eq.~(\protect\ref{ParamV}). The experimental data are from
Refs.~\protect\cite{Exp73} (crosses), \protect\cite{Exp76} (diamonds) and
\protect\cite{Exp78} (circles).\label{lgq}}
\end{figure}

Collectively these results indicate that the phenomenological DSE approach
employed herein, which describes the pion as a bound state of dressed quarks
interacting via nonperturbative gluon exchange, provides a concise, uniformly
good description of the properties of the pion.  This admits the
interpretation that, away from resonances, the nonperturbative dressed-quark
core of the pion is its dominant determining characteristic, as argued in
Refs.~\cite{ABR95,RCSI94}.  

\section{Asymptotic behaviour}
\label{uvform}
A simple way to analyse the behaviour of $F_\pi(q^2)$ at large
spacelike-$q^2$ is to rewrite Eq.~(\ref{LFpi}) in terms of the Bethe-Salpeter
wave-function defined in Eq.~(\ref{BSEwf}):
\begin{eqnarray}
\label{chiF}
\lefteqn{\Lambda_\mu(P+q,-P)= }\\
& &  \frac{2N_c}{f_\pi^2}\,
\int\sfrac{d^4k}{(2\pi)^4}\, {\rm tr}_D
\left[ \overline{\chi}_\pi(k;P+q)
        i\Gamma_\mu(k_{++},k_{-+})
        \chi_\pi(k-\sfrac{1}{2}q;-P) S^{-1}(k_{--})\right]~.\nonumber
\end{eqnarray}

Asymptotic freedom, which is represented up to $\ln[p^2]$-corrections by
Eqs.~(\ref{SUV}) and (\ref{GUV}), entails that the behaviour of $F_\pi(q^2)$
at large spacelike-$q^2$ can be obtained from Eq.~(\ref{chiF}) with
\begin{eqnarray}
\label{UVEst}
S(p) \approx \frac{1}{i\gamma\cdot p + M_c}\;\; & \;\; \mbox{and}\;\;& \;\;
\chi_\pi(p;P) \approx \gamma_5\,
        \frac{\Lambda_1^3}{(p^2+\Lambda^2_2)\,(p^2+M_c^2)}~,
\end{eqnarray}
where $\Lambda_1$, $\Lambda_2$ and $M_c$ are characteristic parameters [the
value of which is unimportant but typically~\cite{ABR95} $\Lambda_1\sim
500$~MeV~$\sim\Lambda_2$ and $M_c \sim 220$~MeV] and $\Gamma_\mu(p_1,p_2) \approx
\gamma_\mu$.  It follows from Eqs.~(\ref{LamFQ}), (\ref{chiF}) and
(\ref{UVEst}) that, at large spacelike-$q^2$,
\begin{equation}
F_\pi(q^2) \propto \int\,\sfrac{d^4 k}{(2\pi)^4}
        \,\chi_\pi(k)\,\chi_\pi(k-\sfrac{1}{2}q)~.
\end{equation}

Making use of the approximation
\begin{eqnarray}
\lefteqn{\sfrac{2}{\pi}
\int_0^{\pi}\,d\theta\,\sin^2\theta\,\frac{f(k^2+p^2 - 2 k p \cos\theta)}
        {k^2+p^2- 2 k p \cos\theta}}\\
&&\approx \frac{f(k^2)}{k^2}\,\theta(k^2 - p^2) 
        + \frac{f(p^2)}{p^2} \,\theta(p^2-k^2)~,
\end{eqnarray}
which is often used in the analysis of the asymptotic behaviour of
Eq.~(\ref{fullDSE}) and is very good for large spacelike-$q^2$~\cite{RM90},
one obtains 
\begin{equation}
\left(\frac{ F_\pi'(x)}{\chi_\pi'(x)}\right)^{'}
 \propto x \,\chi_\pi(x),
\end{equation}
where $x=q^2/4$.  The solution of this equation, which satisfies the boundary
condition \mbox{$F_\pi(q^2=\infty) = 0$}, is
\begin{equation}
\label{Fde}
F_\pi(x) \propto C_1 \, \chi_\pi(x) 
        + \int^x dy\,\chi_\pi'(y)\,\int^y\, dz\, z \,\chi_\pi(z)~,
\end{equation}
where $C_1$ is an undetermined constant. 

One observes from Eq.~(\ref{Fde}) that in impulse approximation the
asymptotic form of the elastic form factor depends on the Bethe-Salpeter
amplitude of the bound state.  This result indicates the failure of the
factorisation Ansatz in the present analysis of this exclusive process.  It
follows because the confinement mechanism explored herein eliminates endpoint
and pinch singularities in Eq.~(\ref{LFpi}).

A dependence of the asymptotic fall-off of $F_\pi(q^2)$ on the pion's
Bethe-Salpeter amplitude is also found using the light-front formulation of
relativistic quantum mechanics~\cite{T76}.  In this approach the asymptotic
form of $F_\pi(q^2)$ is only independent of $\chi_\pi$ if the
constituent-mass of the quark is zero, when the mass-shell singularity
dominates the integral that arises.  In such an approach, however, a
constituent-quark mass of zero is a phenomenologically untenable
assumption~\cite{CCP88}, with a value of $\sim 210$~MeV being required to fit
the available data.

Using the form of $\chi_\pi$ given in Eq.~(\ref{UVEst}), which is a
consequence only of asymptotic freedom, one finds
\begin{equation}
\label{UVFpi} 
F_\pi(q^2) \; \stackrel{q^2\rightarrow \infty}{\propto} \; \frac{\ln q^2}{q^4}~.
\end{equation}
Taking into account the $\ln[p^2]$-corrections to Eqs.~(\ref{SUV}) and
(\ref{GUV}), which arise because of the anomalous dimension of the propagator
and Bethe-Salpeter amplitude in QCD, would only lead to the modification
\mbox{$\ln[q^2] \rightarrow \ln[q^2]^\gamma$}, where $|\gamma|$ is O(1), in
Eq.~(\ref{UVFpi}).  

The numerical methods used herein to calculate $F_\pi(q^2)$ have been
constructed so as to ensure that the result is independent of the details of
the numerical procedure for $0 < q^2 < 20$~GeV$^2$.  Therefore, in order to
verify the result of Eq.~(\ref{UVFpi}) and to estimate the spacelike-$q^2$ at
which the asymptotic regime of this calculation is reached, a least squares
fit of the calculated results for $1/F_\pi(q^2)$ to
\begin{equation}
\label{uvfit}
a_0 + a_1\,\frac{X}{\ln X} + a_2\,\frac{X^2}{\ln X}~,
\end{equation}
with $X = q^2/(1 \;\mbox{GeV}^2)$, was performed on the domain $5 < X <
20$.  This procedure yielded
\begin{equation}
\begin{array}{ccc}
a_0 = 39.2~, & a_1 = -\,14.7~, & a_2 = 1.55~.
\end{array}
\end{equation}
The fitting function and the calculated results are compared in
Fig.~\ref{figUVF}, which confirms Eq.~(\ref{UVFpi}).  This analysis suggests
that, with the parameter values of Eq.~(\ref{ParamV}), which are fixed by
physics at spacelike-$q^2 \leq 4$~GeV$^2$, the asymptotic term; i.e., the
$\ln[X]/X^2$ term, only becomes dominant (provides more than $60$\% of the
magnitude of the form factor) for spacelike-$q^2\gsim 10$~GeV$^2$.  This
result is consistent with the arguments of Ref.~\cite{ILS}; i.e., soft,
nonperturbative physics dominates at presently accessible spacelike-$q^2$ in
exclusive processes.
\begin{figure}
  \centering{\
     \epsfig{figure=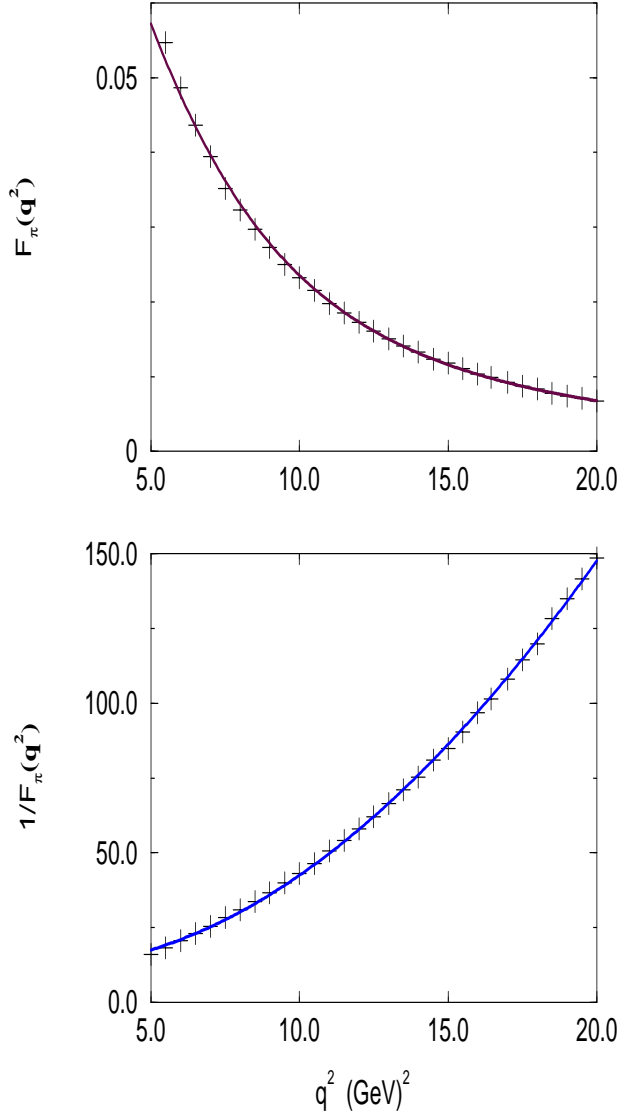,height=16cm,rheight=15cm,width=14cm,angle=-90}  }
\caption{The upper panel compares $F_\pi(q^2)$ with the reciprocal of the
fitting function.  The $a_2$ term provides more than 60\% of the value of
$F_\pi(q^2)$ only for $q^2 > 10$~GeV$^2$.  The lower panel compares
$1/F_\pi(q^2)$ (plotted points) with the fitting function of
Eq.~(\protect\ref{uvfit}).
\label{figUVF} }
\end{figure}

Equation~(\ref{Fde}) follows because of the constraints that the realisation
of confinement explored herein place on the fermion propagator.  This
mechanism is not an essential element of the DSE framework.  The invalidity
of Eq.~(\ref{Fde}), and hence Eq.~(\ref{UVFpi}), would not compromise the
quantitative results discussed in the preceding sections.

\section{Summary and Conclusions}
The impulse approximation, illustrated in Fig.~\ref{figppA}, has been used to
calculate $F_\pi(q^2)$ at spacelike-$q^2$; i.e., away from resonance
contributions.  The important elements of this calculation are: the dressed
quark propagator; the pion Bethe-Salpeter amplitude; and the dressed
quark-photon vertex, which follow from realistic, nonperturbative
Dyson-Schwinger equation studies.  This entails that in this calculation each
of these elements behaves at large spacelike-$p^2$ in the manner prescribed
by the renormalisation group in QCD, up to $\ln[p^2]$-corrections, and at
small spacelike-$p^2$ in such a way as to ensure confinement.  In addition,
the same approach has been used to simultaneously calculate: $f_\pi$;
$m_\pi$; $\langle \bar q q \rangle$; $r_\pi$; the \mbox{$\pi^0\rightarrow
\gamma\gamma$} decay width; the $\pi$-$\pi$ scattering lengths: $a_0^0$,
$a_0^2$, $a_1^1$, $a_2^0$, $a_2^2$; and the associated partial wave
amplitudes.

The calculated results for all quantities agree well with the data, as
discussed in Sec.~\ref{secivb}.  This supports the contention that the
confined, nonperturbative ``dressed-quark core'' is the dominant determining
characteristic of the pion, away from resonance contributions.  Particularly
interesting in this context is the calculation of the $\pi^0\rightarrow
\gamma\gamma$ decay width, Sec.~\ref{secpigg}.  The correct chiral limit
value for this decay width is obtained {\it independent} of the details of
the modelling of the quark propagator.  This is an illustration of the manner
in which the Abelian anomaly manifests itself in the Dyson-Schwinger equation
approach.  This calculation indicates that soft, nonperturbative
contributions dominate the exclusive electromagnetic form factor for
spacelike-$q^2\lsim 10$~GeV$^2$.

The Dyson-Schwinger equation approach provides a phenomenological framework
in which to relate experimental observables to the qualitative and
quantitative features of the effective quark-quark interaction in the
infrared, which is an important but presently unknown aspect of QCD.  There
have been attempts to calculate the gluon propagator directly: some by
solving the Dyson-Schwinger equation for the gluon vacuum
polarisation~\cite{Baker,Atkin,Pennington}; and others via lattice
simulations~\cite{BPS94}, which are presently qualitatively and
quantitatively unreliable for spacelike-$q^2\lsim 1$~GeV$^2$.  The present
study suggests that the framework explored herein can be a valuable
complement to such analyses.  More precise measurements of $F_\pi(q^2)$ for
spacelike-$q^2\gsim 1$~GeV$^2$ would be useful in this regard.

{\bf Acknowledgments}.\\
This work was supported by the US Department of Energy, Nuclear Physics
Division, under contract number W-31-109-ENG-38.  The calculations described
herein were carried out using a grant of computer time and the resources of
the National Energy Research Supercomputer Center.

%______________________________ References ______________________________

%____________________________________________________________________________
\end{document}